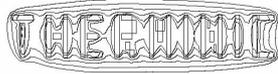

# Application of Structure Functions for the Investigation of Forced Air Cooling

Marcin Janicki *, Jedrzej Banaszczyk #, Gilbert De Mey #, Marek Kaminski *,
Bjorn Vermeersch #, Andrzej Napieralski *
* Department of Microelectronics and Computer Science, Technical University of Lodz,
Al. Politechniki 11, 93-590 Lodz, Poland
# Department of Electronics and Information Systems, University of Ghent,
Sint Pietersnieuwstraat 41, 9000 Ghent, Belgium

*Abstract-* This paper presents thermal analyses of a power amplifier placed in a wind tunnel. All the investigations are based on the transient temperature measurements performed during the circuit cooling process. The measured cooling curves were used to compute the cumulative and differential structure functions for the circuit with a heat sink. These functions helped to determine the optimal values of circuit model parameters necessary for numerical thermal simulations. The experiments demonstrated the influence of the wind speed on the value of the heat transfer coefficient and consequently on the temperature of the entire structure.

I. INTRODUCTION

One of the main problems in dynamic thermal simulations of electronic circuits consists in the determination of thermal model parameters, such as thermal conductivity and capacity or the heat transfer coefficient. Some of these parameters can be found based on adequate transient thermal measurements. This paper will demonstrate how advanced thermal analysis tools, i.e. the structure functions, can be employed for the optimisation of circuit thermal models.

The next section of this paper presents briefly the theory of circuit transient temperature measurements and dynamic thermal analysis tools. Next, the theory is employed in the experimental part of the paper to investigate, based on the example of a power amplifier, the effects of variable forced air cooling. Finally, the results of thermal analyses are used in the optimisation of a circuit thermal model.

II. ANALYSIS OF TRANSIENT THERMAL STATES

Before thermal analyses can be carried out it is necessary to perform adequate temperature measurements of a given structure. The temperature measurements of an electronic circuit can be taken using a forward biased p-n junction. When the bias current is constant, the voltage drop across the junction serves as the measure of temperature. Theoretically, circuit temperature should be measured as a response to the power step excitation. However, this solution requires using the same junction both for heating and measurement, which might cause technical problems. An alternative approach, used also in this paper, is to heat the junction till steady state condition is reached and then to switch the power off and capture a cooling curve. Then, when nonlinear effects are negligible, this curve is just the complement of the heating curve and contains the same information.

The most important issue in the measurements of thermal transient responses is the time resolution of the recorded data. Namely, measured structures consist of multiple layers made of different materials (silicon chip, package, cooling assemblies), each having different geometry and thermal properties. Thus, the thermal responses are a superposition of many exponential curves corresponding to different time constants, which might span even over several decades. Therefore, equidistant sampling on the logarithmic time scale should be used for the acquisition of the thermal transient curves. Only then all the time constants can be identified. Similarly, because the thermal transients are very rapid, the most important temperature value changes happen in the very beginning. Consequently, the logarithmic time scale should be used also when presenting graphically results of transient thermal measurements.

The thermal time constant spectrum can be identified from the measurements, e.g. employing the Network Identification by Deconvolution (NID) method; originally developed in the eighties by Szekely and van Bien [1]. According to this method, the entire time-constant spectrum is computed from the equally spaced, on the logarithmic time scale, thermal transient data computing firstly the time derivative of the captured curve and then performing a deconvolution.

The computed time-constant-spectra can be used to obtain the so-called structure functions, which are extremely useful for the thermal analysis of electronic circuits. Namely, with limited accuracy, the continuous spectra can be discretized, thus obtaining the Foster RC ladder canonical form. Because this ladder does not contain thermal capacitances connected to ground, it is physically erroneous and has to be converted to the Cauer ladder canonical form, so that to properly reflect thermal processes within a structure.

The Cauer RC networks can be represented, as illustrated in Figure 1, by the cumulative structure functions $C_\Sigma(R_\Sigma)$, constituting a kind of thermal resistance and capacitance map for the entire heat-flow path. The origin corresponds to the area where power is dissipated, while the singularity at the end can be associated with the ambient. The plateaus in the curve are related to a certain mass of material from where the corresponding thermal capacitance values, $C_{th}$, can be read.

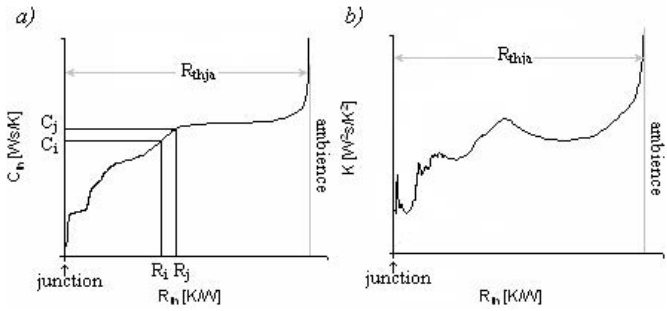

Fig. 1. Exemplary cumulative (a) and differential (b) structure functions.

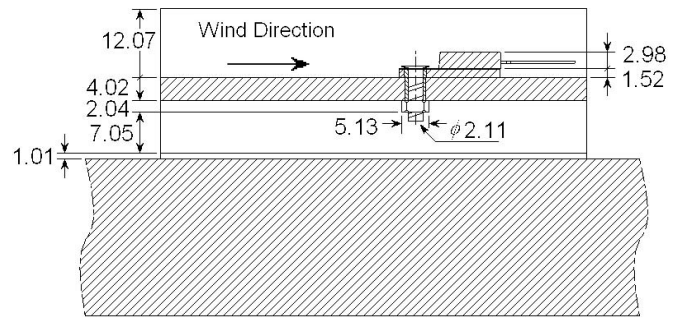

Fig. 3. Side view of the amplifier with the radiator.

The derivative of the cumulative thermal capacitance with respect to the thermal resistance is the differential structure function. This function, also shown in Figure 1, is extremely useful as well. Namely, the peaks in this curve correspond to the change of a material through which heat is diffusing, so they can be attributed accordingly to the chip, the package and the cooling assembly [2].

III. MEASUREMENTS

A. Measurement Setup

For the experiments the APEX PA60 power amplifier was used. This circuit was chosen expressly to simplify the entire measurement procedure owing to the fact that the package contains twin power operational amplifiers with their outputs protected by diodes connected between the output and the positive power supply. In such a case, one of the amplifiers was powered in order to provide the heat dissipation and the protection diode of the other one was used as the temperature sensor. Because the temperature measurement point does not coincide exactly with the location of the heat source, there was a slight delay of 30 µs observable between the sensed temperature and the temperature of the heat source.

The power amplifier, as shown in Figs. 2-3, was screwed to an aluminium heat sink. Additionally, thermal grease was introduced between the package and the radiator to reduce the thermal resistance. The total outer surface of the sink was equal to 480 cm$^2$. All the dimensions in the figures are given in millimetres.

The whole assembly was placed in the horizontal position inside the wind tunnel, as shown in Figure 4, on a thermal insulator about 15 cm from the bottom of the tunnel (about one third of its height). In this region the air velocity profile is known to be uniform and the air flow is laminar. The wind direction is indicated in Figures 2-4 by arrows.

B. Wind Speed Influence

The measurements were performed using the thermal tester T3Ster manufactured by the MicRed company. The heating amplifier was operated in the saturation mode with the power dissipation of 10 W till steady state temperature was reached. After switching the power off, the protection diode of the other amplifier was used to measure the cooling curve. The forward current through the diode during the measurement was 1 mA. The measured cooling curves for the air velocities varying between 0 m/s (still air) and 4.15 m/s (maximal air speed) are shown in Figure 5. These curves were processed further using the software provided by the MicRed company together with the thermal tester. As a result, the cumulative and differential structure functions, presented in Figures 6-7 respectively, were obtained.

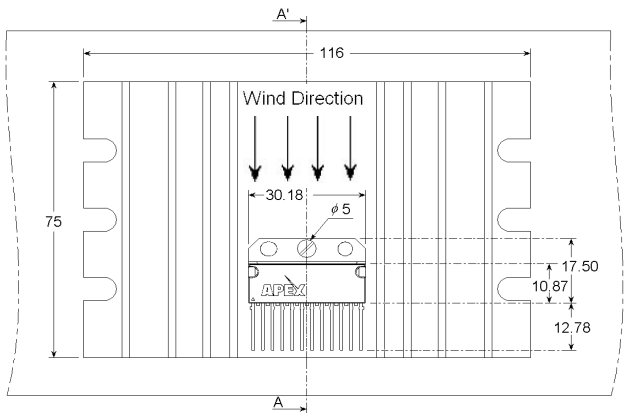

Fig. 2. Top view of the amplifier with the radiator.

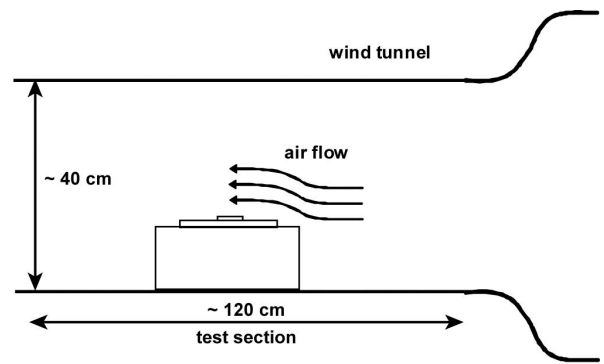

Fig. 4. Side view of the wind tunnel.

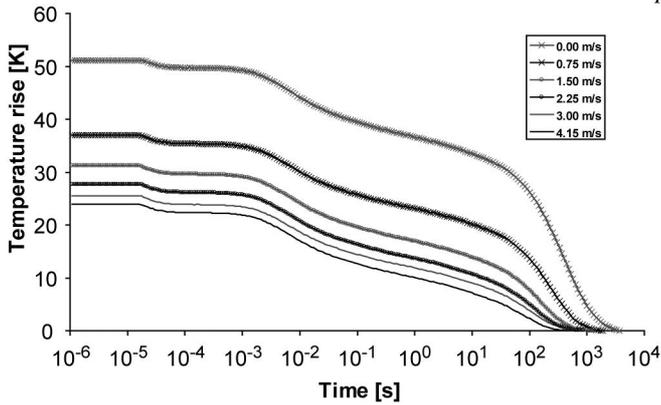

Fig. 5. Cooling curves for variable wind speeds.

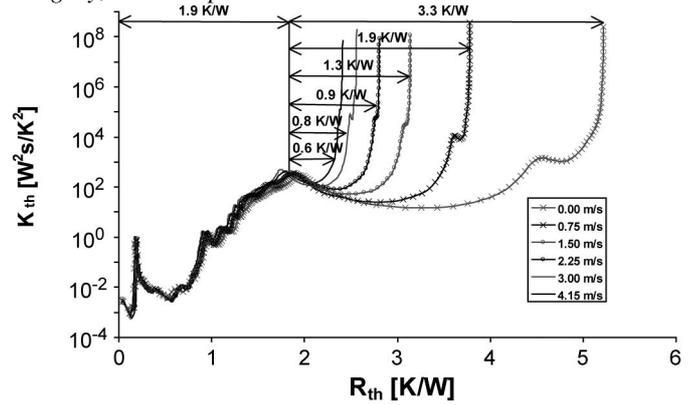

Fig. 7. Differential structure functions.

Looking at the structure functions it can be concluded that they differ only at their sections corresponding to the cooling fin. This result could be expected because the variation of the air velocity affects only the heat exchange with the ambient whereas the heat flow within the package is not influenced at all. As can be estimated from the figures, owing to the forced cooling the total thermal resistance can be more than halved from over 5.2 K/W to 2.4 K/W.

Two plateaus encircled in Figure 6 can be attributed to the silicon die and the heat sink [3]. Theoretically, as mentioned previously, the heat capacity values at these plateaus should render possible the estimation of the semiconductor chip and radiator volumes. The cumulative capacitance value of the second plateau is 130 J/K, which for the aluminium material data (density and specific heat) yields the estimated radiator volume of 53 cm$^3$, whereas the real one is 50 cm$^3$. Taking into account that the cumulative capacitance value includes also the package, the estimation result seems very accurate. On the other hand, the estimated volume of the silicon chip is 6 mm$^3$, whereas the measured one is only 1.7 mm$^3$. This discrepancy can be explained by the fact that the silicon chip is attached to the 1.5 mm thick metal plate by an irregular solder layer, whose thickness reaches almost 1 mm. Thus, the cumulative capacitance of the first plateau corresponds rather to the total capacitance of the chip with the die attach.

Because the forced cooling influences the heat exchange with the ambient only at outer surfaces of the structure, all the computed curves are overlaid in the areas corresponding to the inner parts of the package. Then, they differ slightly in the section where the heat flows from the package to the heat sink to diverge finally, approximately at a thermal resistance of 1.9 K/W.

Thus, knowing the radiator surface, the thermal resistances indicated in Figure 7 may be used to determine the average values of the total heat transfer coefficient. The heat transfer coefficient value for the radiation with the natural convection (the air speed of 0m/s) was found to be equal to 6.3 W/m$^2$K, which is almost exactly the value obtained from the empirical formula provided in [4] for the heat sink temperature rise of 32 K at the ambient temperature of 300 K.

Then, the heat transfer coefficient values due to the forced cooling, presented in Figure 8, were computed by subtracting the coefficient value determined without the forced cooling (this value should not change too much because, according to the ratios of thermal resistances, the radiator temperature rise decreased from only 32 K to 6 K at the full wind speed) from the total heat transfer coefficient. Additionally, as suggested in [5], these values were fitted as an exponential function of the wind speed. The best fit was obtained for nearly linear dependence of the heat transfer coefficient on the air speed.

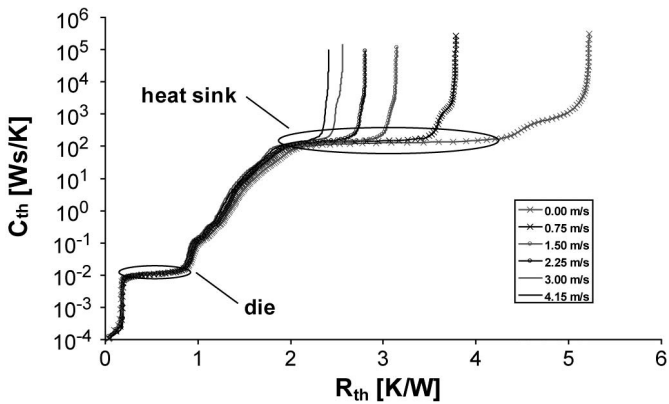

Fig. 6. Cumulative structure function.

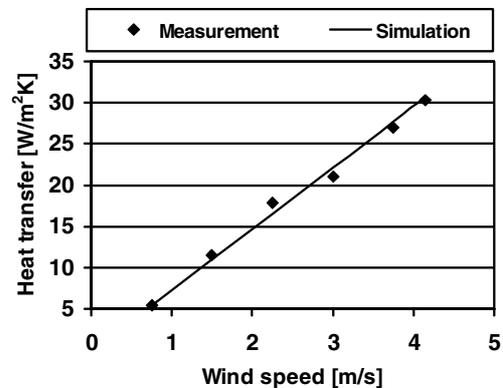

Fig. 8. Average forced convection heat transfer coefficient vs. wind speed.

TABLE I
SIMULATION DATA.

| Material | Number of blocks [-] | Number of nodes [-] | Thermal conductivity [W/mK] | Thermal capacity [J/m$^3$K] |
|---|---|---|---|---|
| Silicon | 1 | 72 | 102.0 | $5.50*10^6$ |
| Copper | 14 | 585 | 373.0 | $3.48*10^6$ |
| Plastic package | 7 | 724 | 0.153 | $2.30*10^6$ |
| Aluminium radiator | 20 | 1557 | 164.0 | $2.43*10^6$ |
| Σ | 42 | 2938 | | |

## IV. THERMAL SIMULATIONS

The data extracted based on the transient measurements were used then for the thermal simulations of the structure. For this purpose, the detailed three-dimensional model was created. The structure was discretized and the sensing diode temperature was computed with the finite difference solver TULSOFT. The initial thermal parameter values used in the model were optimised to match the transient measurements. The obtained simulation results for two extreme cases are compared with the measurements in Figures 9-10 and the final parameter values together with the information on the structure discretisation mesh are given in Table 1.

As can be seen in the figures, the simulation results agree with the measurements quite accurately. Moreover, all the thermal data have reasonable values, except for the thermal capacity of silicon, which is three times higher than the real one. This however, as already discussed, is caused the most probably by the presence of the thick die attach layer.

## V. CONCLUSIONS

This paper demonstrated in practice the usefulness of the structure functions in the determination of different thermal model parameters. In particular, the transient measurements allowed the identification of individual materials and their volumes. Moreover, knowing the geometry of the structure it was possible to determine the average values of the heat transfer coefficient in all the cases. Owing to the proposed approach, it was possible to optimize the thermal models and increase greatly the dynamic simulation accuracy.

### ACKNOWLEDGEMENTS

Bjorn Vermeersch, a Research Assistant for the Scientific Research Foundation - Flanders, would like to thank them for their financial support.

This research was supported partly by the grant of the Polish Ministry of Science and Higher Education No. N515 008 31/0331.

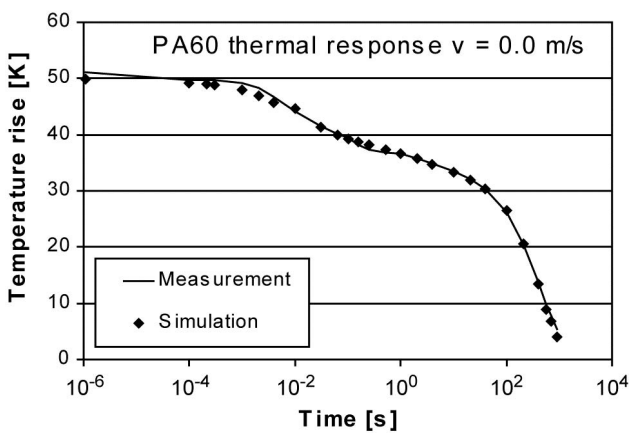

Fig. 9. Measured and simulated cooling curves - wind speed 0 m/s.

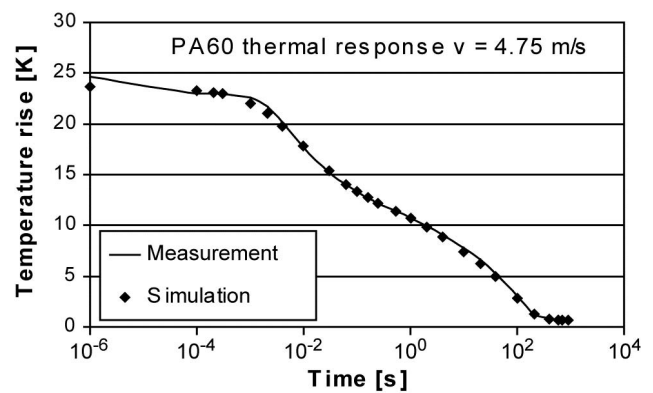

Fig. 10. Measured and simulated cooling curves - wind speed 4.15 m/s.